\documentstyle[12pt]{article}

\begin{document}

\title{The quantum modes of the (1+1)-dimensional 
       oscillators in  general relativity}

\author{Ion I. Cot\u aescu\\ 
and\\
Ion I. Cot\u aescu Jr.\\
{\small \it The West 
University of Timi\c soara,}\\ {\small \it V. P\^ arvan Ave. 4, RO-1900 
Timi\c soara, Romania}}

\date{\today}

\maketitle

\begin{abstract}
The quantum modes of a new family of relativistic oscillators are studied 
by using the supersymmetry and  shape invariance in a version suitable 
for (1+1) dimensional relativistic systems. In this way one obtains the 
Rodrigues formulas of the normalized energy eigenfunctions of the discrete 
spectra and the corresponding rising and lowering operators.    

Pacs:  04.62.+v, 03.65.Ge
\end{abstract}
\

\newpage

\section{Introduction}
\

In general relativity, the geometric models play the role of  
kinetics, helping us to understand the characteristics of the  classical or 
quantum free motion on a given background. One of the simplest geometric 
models in (1+1) dimensions is that of the quantum relativistic  
oscillator (RO) defined as a free massive scalar particle  on the  anti-de 
Sitter static background \cite{BRD,OSC,N}.  Recently, we have generalized this 
model to a  family of quantum models of RO whose metrics are one-parameter 
deformations (i.e. conformal transformations) of the anti-de Sitter or de 
Sitter ones \cite{CO1}. As it is shown in Refs.\cite{CO2}, the deformed anti-de 
Sitter metrics give the  relativistic correspondents the usual  nonrelativistic 
P\" oschl-Teller (PT) problems while the deformed de Sitter metrics generate 
relativistic  Rosen-Morse (RM) problems \cite{PT}. A remarkable  property of  
these RO is that  all of them have as nonrelativistic limit just the  usual 
nonrelativistic harmonic oscillator (NRHO) \cite{CO1}.

In  these relativistic models the Klein-Gordon equation is analytically 
solvable in the same manner as the Schr\" odinger equation of the 
mentioned well-studied nonrelativistic problems. This allows one to study the 
RO by using the successful methods of supersymmetry and shape 
invariance \cite{SUP} with minimal changes requested by the specific form the 
Klein-Gordon equation \cite{CA1}. In this way one can derive the normalized 
energy eigenfunctions of the discrete energy spectrum and the form of the 
shift operators of the energy basis that are involved in the structure of the 
dynamical algebras \cite{CA2}.

Here we would like to present a systematic study of our family of RO based 
on the  supersymmetry and shape invariance of the relativistic potentials, 
pointing out the main specific features of the PT and RM relativistic problems.  
We believe that  this first example of a family of metrics generating  
analytically solvable quantum problems could be of interest for further 
investigations concerning the supersymmetry of other solvable relativistic 
quantum models in (3+1) dimensions \cite{AIS,CO3} or more \cite{CO4}.

We start in Sec.2 with a short review of the (1+1) relativistic scalar quantum 
mechanics constructed as the one-particle restriction of the theory of 
the scalar free field on curved space-time. We define the state space in the 
coordinate representation and we introduce the coordinate 
and momentum operators. In Sec.3 we present the  relativistic PT and RM 
oscillators giving their energy spectra and the energy eigenfunctions up to 
normalization factors. The relativistic supersymmetry and the shape invariance 
of the relativistic PT and RM potentials are used in the next section for 
deriving the definitive form of the normalized energy eigenfunctions of the 
discrete energy spectra. The Sec.5. is devoted to the properties of the 
shift operators of the energy bases of our RO. Therein we recover the 
known shift operators of the PT models \cite{PT1} and we derive those of 
the RM models.

\section{Relativistic quantum mechanics}
\

It is well-known that the one-particle relativistic quantum mechanics cannot 
be constructed as an  independent consistent theory because of some 
difficulties related to the probabilistic interpretation of  the relativistic 
wave functions. The good theory of the  relativistic quantum 
systems is in fact the quantum field theory where the second quantization 
guarantees a coherent probabilistic interpretation. In these conditions, 
the relativistic quantum mechanics, in the sense of general relativity, can be 
seen as the one-particle restriction of the quantum field theory  on curved 
backgrounds. Herein the quantum modes of the scalar particle in external 
gravitational field are given by the particular solutions of the free 
Klein-Gordon  equation. These may form a basis in the space of wave functions  
organized as a Hilbert space with respect to the scalar product derived from 
the expression of the conserved electric charge \cite{BD}.

\subsection{The Klein-Gordon equation in special frames} 
\

Let us consider a (1+1)-dimensional background with a static local chart 
(i.e., natural frame) of holonomic 
coordinates $(u^{0},u^{1})\equiv (t,u)$ where the metric tensor defined on 
the space domain $D_{u}$  
is $g_{\mu \nu}(u)$,  $\mu, \nu =0,1$, and  $g={\rm det}(g_{\mu \nu})$. 
The one-particle quantum modes of a scalar field $\phi$ of the mass $m$, 
minimally coupled with the gravitational 
field, are given by the  Klein-Gordon equation  
\begin{equation}\label{(kg)}
\frac{1}{\sqrt{-g}}\partial_{\mu}\left(\sqrt{-g}g^{\mu\nu}\partial_{\nu}\phi
\right) + m^{2}\phi=0\,,
\end{equation}
written in natural units with $\hbar=c=1$.
Since in the static charts the energy, $E$, is conserved,  the Klein-Gordon 
equation has a set of fundamental solutions (of positive and 
negative frequencies), 
\begin{equation}\label{(sol)}
\phi_{E}^{(+)}(t,u)=\frac{1}{\sqrt{2E}}e^{-iEt}U_{E}(u)\,, \qquad 
\phi^{(-)}=(\phi^{(+)})^{*}\,,
\end{equation}
which depend on the static energy eigenfunctions $U_{E}$. These may be 
orthonormal (in usual or generalized sense) with respect to the relativistic 
scalar product \cite{BD} 
\begin{equation}\label{(psc1)}
\left<U,U'\right>=\int_{D_{u}}du~\mu(u)U(u)^{*}U'(u)
\end{equation}
where
\begin{equation}\label{(mu)}
\mu=\sqrt{-g}\,g^{00}
\end{equation}
is the specific relativistic weight function of the scalar field. The Hilbert 
space of the square integrable functions with respect to this scalar product 
is denoted by ${\cal L}^{2}(D_{u},\mu)$. Oviously, the set of the wave 
functions $U_{E}$ represents an usual or generalized basis in this space. This  
will be called the energy basis.

In the case of the  static backgrounds, a change of the space coordinates does 
not change the quantum modes. On the other hand, it is known that in (1+1) 
dimensions any  static background has a {\em special} natural frame where the 
metric is the conformal transformation of the Minkowski flat one.  
Starting with any natural frame $(t,u)$, the space coordinate of the special 
frame $(t,x)$ reads  
\begin{equation}\label{(chi)}
x=\chi (u) = \int du\,\mu (u) + {\rm const.}\,, 
\end{equation}
where the constant assures the condition $\chi (0)=0$.  
In the special frame 
we have $\tilde g_{00}(x)=-\tilde g_{11}(x)$ and $\tilde\mu(x)=1$. 
Therefore, the scalar product (\ref{(psc1)}) becomes just the usual one 
of the Hilbert space ${\cal L}^{2}(D)$ where $D$ is the  domain of the 
coordinate $x$  corresponding to $D_{u}$.

If we denote   $\tilde g_{00}=1+v$, then we can write the line element 
of the special frame as 
\begin{equation}\label{(met)}
ds^{2}=[1+v(x)](dt^{2}-dx^{2})
\end{equation}
while  the Klein-Gordon equation takes the form
\begin{equation}\label{(kkg)}
\left[-\frac{d^2}{dx^2}+V_{R}(x)\right]U_{E}(x)=(E^{2}-m^{2})U_{E}(x)
\end{equation}
where  $V_{R}=m^{2}v$. We say that this is the relativistic potential since in  
the nonrelativistic limit $V_{R}/2m$ becomes just the usual  
potential of the corresponding Schr\" odinger equation.

\subsection{Observables}
\

The linear operators on ${\cal L}^{2}$  (denoted here using boldface)  
can be defined either by giving their matrix elements in  a countable 
basis  of ${\cal L}^{2}$ or as differential operators in the coordinate 
representation.
The most general  differential operator we  use in an arbitrary 
frame $(t,u)$ has the form
\begin{equation}\label{(opd)}
({\bf D}U)(u)=i\left[f(u)\frac{d}{du}+h(u)\right]U(u)\,,
\end{equation}
depending on two  real functions $f$ and $h$.
Its  adjoint with respect to the scalar product (\ref{(psc1)}) is
\begin{equation}\label{(opda)}
{\bf D}^{\dagger}={\bf D}+i\left[\frac{1}{\mu}\frac{d(\mu f)}{du}-2h\right]
{\bf 1}
\end{equation}  
where  ${\bf 1}$ is the identity operator. 
Hereby, we see that for $h=\partial_{u}(\mu f)/2\mu$ the operator ${\bf D}$ is 
self-adjoint \cite{DW,CA2}.

The consequence is that we can introduce a unique pair of self-adjoint 
coordinate and momentum operators defining their action in any natural frame 
$(t,u)$. One can easily verify that the following definitions,
\begin{equation}
({\bf X}U)(u)=\chi(u)U(u)\,, \qquad 
({\bf P}U)(u)=i\frac{1}{\mu(u)}\frac{dU(u)}{du}\,,  
\end{equation}
are satisfactory since the commutation relation  
\begin{equation}
[{\bf P},{\bf X}]=i{\bf 1} 
\end{equation}
is just the desired one. Obviously, in the special frame $(t,x)$ these 
operators have the same action as in the case of the Minkowski flat space-time, 
namely
\begin{equation}
({\bf X}U)(x)=xU(x), \qquad ({\bf P}U)(x)=i\frac{dU(x)}{dx}\,.  
\end{equation}
Furthermore, one can put the Klein-Gordon equation (\ref{(kkg)}) in operator 
form,
\begin{equation}
{\bf H}^{2}=m^{2}{\bf 1}+{\bf \Delta}[V_{R}]\,, 
\end{equation}
where  ${\bf H}$ is the Hamiltonian operator defined by  
${\bf H}U_{E}=EU_{E}$ and 
\begin{equation}\label{(Delta)}
{\bf \Delta}[V]={\bf P}^2 + V({\bf X})\,.
\end{equation}

Thus we have obtained the main operators on ${\cal L}^{2}$. The whole algebra 
of observables is  that freely generated by the operators ${\bf X}$ and 
${\bf P}$, like in the Schr\" odinger picture of the nonrelativistic 
one-dimensional quantum mechanics.

\section{Relativistic Oscillators}
\

The new geometric models of RO we discuss here are simple systems 
of free test particles that move on static backgrounds simulating oscillations. 
This means that 
there are local charts of coordinates $(t,u)$ where an observer at $u=0$ moving 
along the direction $\partial_{t}$ observes an oscillatory geodesic motion.  
These charts called {\em proper} natural frames have line elements \cite{CO1},
\begin{equation}\label{(m)}
ds^{2}=g_{00}dt^{2}+g_{11}du^{2}=
\frac{1+(1+\lambda) \omega^{2}u^{2}}{1+\lambda \omega^{2} u^{2}}dt^{2}-  
\frac{1+(1+\lambda) \omega^{2}u^{2}}{(1+\lambda \omega^{2} u^{2})^{2}}du^{2}\,,
\end{equation} 
depending on a real parameter $\lambda$. Thus one obtains a family of metrics 
which are conformal transformations either of the anti-de Sitter metric 
(as given in Ref.\cite{BRD}) or of the de Sitter one. The anti-de Sitter metric 
with  $\lambda = -1$ is also included in this family. A special case is that of 
$\lambda=0$ when we say that the line element
\begin{equation}\label{(mno)} 
ds^{2}=(1+\omega^{2}u^{2})(dt^{2}-du^{2})
\end{equation}
defines the {\em normal} RO. In Ref \cite{CO1} it is shown that the quantum 
models with $\lambda\le 0$ have countable energy  spectra  while for 
$\lambda>0$ the energy spectra are mixed, with a finite discrete sequence 
and a continuous part. All these models will be presented here in the special 
frames $(t,x)$ associated  with the proper frames $(t,u)$ defined  above. 
The advantage is that in the special frames our RO  appear either as PT or as 
RM relativistic systems \cite{CO2} which can be analytically solved like 
those known from the nonrelativistic quantum mechanics.  

\subsection{The relativistic P\" oschl-Teller models}
\

Let us  consider first the models with $\lambda<0$ when the metrics are 
conformal transformations of the anti-de Sitter one. We denote 
\begin{equation}\label{(parpt)}
\lambda=-\epsilon^{2}, \qquad \hat\omega =\epsilon\, \omega, \qquad 
\epsilon\ge 0
\end{equation} 
and calculate  the space coordinate of the special frame. According to 
Eq.(\ref{(chi)}) we obtain 
\begin{equation}\label{(xPT)}
x=\frac{1}{\hat\omega}\arcsin\hat\omega u
\end{equation}
while from Eq.(\ref{(m)}) we write the line element in this frame,    
\begin{equation}\label{(mpt)}
ds^{2}=\left( 1+\frac{1}{\epsilon^2}\tan^{2}\hat\omega  x\right)(dt^{2}
-dx^{2})\,,
\end{equation} 
where the space domain  is  $D =(-\pi/2\hat\omega, \pi/2\hat\omega)$ 
because of  the event horizon at $\pm \pi/2\hat\omega$. 
The relativistic potential,   
\begin{equation}\label{(vpt)}
V_{PT}(k,x)=\frac{m^2}{\epsilon^2}\tan^{2}\hat\omega x =
\hat\omega^{2} k(k-1) \tan^{2}\hat\omega x\,,
\end{equation}
is a PT one depending on the new parameter   
\begin{equation}\label{(kpt)}
k=\sqrt{\frac{m^{2}}{\epsilon^{2}\hat\omega^{2}}+\frac{1}{4}}+\frac{1}{2}\,.  
\end{equation}
In the following we use $k$ instead of $m$, as the  main parameter of 
the PT models that will be denoted from now by $(k)$.  

The Klein-Gordon equation (\ref{(kkg)}) of the model $(k)$ with the 
potential (\ref{(vpt)}) can be written as 
\begin{equation}\label{(kgPT)}
\left[-\frac{1}{\hat\omega^2}\frac{d^2}{dx^2}
+\frac{k(k-1)}{\cos^{2}\hat\omega x}\right]U(x)=\nu^{2} U(x)
\end{equation}
where 
\begin{equation}
\nu^{2}=\frac{E^2}{\hat\omega^2}-\left(1-\frac{1}{\epsilon^2}\right)
\frac{m^2}{\hat\omega^2}=\frac{E^2}{\hat\omega^2}+(1-\epsilon^{2})k(k-1)\,.
\end{equation}
Its solutions
\begin{equation}\label{(solPT)}
U(x)\sim \sin^{2s}\hat\omega x\cos^{2p}\hat\omega x
F\left(s+p-\frac{\nu}{2},
s+p+\frac{\nu}{2}, 2s+\frac{1}{2}, \sin^{2}\hat\omega x\right)\,,
\end{equation}
are expressed in terms of Gauss hypergeometric functions \cite{AS} whose  
parameters $s$ and $p$ are solutions of the equations  $2s(2s-1)=0$ 
and $2p(2p-1)=k(k-1)$.  The wave functions (\ref{(solPT)})  have good physical 
meaning only when  $F$ is a polynomial selected by a suitable 
quantization condition (since otherwise $F$ is strongly divergent for 
$x\to \pm\pi/2\hat\omega$). Therefore, we introduce the quantum number 
$n_{s}$ and impose  
\begin{equation}\label{(quant)}
\nu=2 (n_{s}+s+p)\,,\quad n_{s}=0,1,2,...\,.
\end{equation}
In addition, we choose the boundary conditions of the {\em regular} modes 
\cite{AIS} given by  $2s=0,1$ and $2p=k$. Then the energy levels   
\begin{equation}\label{(labpt)}
{E_{k,n}}^{2}=\hat\omega^{2}[(k+n)^{2}+(\epsilon^{2}-1)k(k-1)] 
\end{equation}
depend only on the main quantum number, $n=2n_{s}+2s$, which takes  even values 
if $s=0$ and odd values for $s=1/2$. Particularly for the anti-de Sitter 
model with $\epsilon=1$ we recover the well-known result 
$E_{k,n}=\omega(k+n)$ \cite{N}.

The next step is to derive the concrete form of the normalized energy 
eigenfunctions corresponding to these energy levels. According to 
Eqs.(\ref{(solPT)}) and (\ref{(quant)}), these are
\begin{equation}\label{(solPT1)}
U_{k,n}(x)=N_{k,n}\sin^{2s}\hat\omega x\cos^{k}\hat\omega x
F\left(-n_{s},
n_{s}+k+2s, 2s+\frac{1}{2}, \sin^{2}\hat\omega x\right)\,,
\end{equation}
where the normalization constants $N_{k,n}$ might be calculated with the help 
of the scalar product of ${\cal L}^{2}(D)$. However, there is another efficient 
method based on supersymmetry and shape invariance \cite{SUP}
giving  directly the Rodrigues formula of the normalized 
eigenfunctions. This will be presented in the next section.  

We specify that our PT models are well-defined for any $k\in [1,\infty)$ since 
the limit $k\to 1$ (when $m\to 0$) has a good physical meaning. Indeed, in this 
case the massless particle remains confined to the rectangular infinite well 
of width $\pi/\hat\omega$ having the equidistant energy levels 
\begin{equation}
E_{1,n}=\hat\omega (n+1),
\end{equation}   
corresponding to the normalized eigenfunctions
\begin{equation}\label{(u1n)}
U_{1,n}(x)= \sqrt{\frac{2\hat\omega}{\pi}}\sin (n+1)\left(\frac{\pi}{2}-
\hat\omega x\right),
 \quad n=0,1,2,... \,.
\end{equation}
Notice that this is a pure relativistic model since its nonrelativistic 
limit does not make sense.

\subsection{The relativistic Rosen-Morse models}
\

For $\lambda>0$ the metrics of RO are conformal transformations of the de 
Sitter metric. Now  we change the significance of $\epsilon$ and  put 
\begin{equation}\label{(parrm)}
\lambda=\epsilon^{2}, \qquad \hat\omega =\epsilon\, \omega, \qquad 
\epsilon\ge 0\,.
\end{equation} 
Furthermore, from  Eq.(\ref{(chi)}) we find  
\begin{equation}\label{(xRM)}
x=\frac{1}{\hat\omega}{\rm arcsinh}\,\hat\omega  u 
\end{equation}
and from Eq.(\ref{(m)}) we obtain  the  line element  
\begin{equation}
ds^{2}=\left( 1+\frac{1}{\epsilon^2}\tanh^{2}\hat\omega  x\right)(dt^{2}
-dx^{2})
\end{equation} 
in the special frame where the space domain is $D =(-\infty, \infty)$. 
These metrics define  relativistic RM models whose potentials,  
\begin{equation}\label{(vrm)}
V_{RM}(j,x)=\frac{m^2}{\epsilon^2}\tanh^{2}\hat\omega x
=\hat\omega^{2}j(j+1) \tanh^{2}\hat\omega x\,,
\end{equation} 
depend on the parameter
\begin{equation}\label{(krm)}
j=\sqrt{\frac{m^{2}}{\epsilon^{2}\hat\omega^{2}}+\frac{1}{4}}-\frac{1}{2}\,.
\end{equation}
Like in the case of PT models, we consider that $j$ is the main parameter of 
the RM models, denoted by $(j)$.

Now the Klein-Gordon equation is  
\begin{equation}\label{(kgRM)}
\left[\frac{1}{\hat\omega^2}\frac{d^2}{dx^2}
+\frac{j(j+1)}{\cosh^{2}\hat\omega x}\right]U(x)=\hat\nu^{2} U(x)
\end{equation}
where 
\begin{equation}
\hat\nu^{2}=-\frac{E^2}{\hat\omega^2}+\left(1+\frac{1}{\epsilon^2}\right)
\frac{m^2}{\hat\omega^2}=-\frac{E^2}{\hat\omega^2}+(1+\epsilon^{2})j(j+1)\,.
\end{equation}
The  solutions
\begin{equation}\label{(solRM)}
U(x)\sim \sinh^{2s}\hat\omega x\cosh^{2p}\hat\omega x
F\left(s+p-\frac{\hat\nu}{2},
s+p+\frac{\hat\nu}{2}, 2s+\frac{1}{2}, -\sinh^{2}\hat\omega x\right)\,,
\end{equation}
depend on the parameters $s$ and $p$ which  satisfy 
$2s=0,1$ and $2p(2p-1)=j(j+1)$.
Like in the nonrelativistic case, the relativistic RM models have  mixed 
energy spectra with a finite discrete sequence and a continuous part 
\cite{CO1}. 

The discrete levels arise from the quantization condition 
\begin{equation}\label{(quant)}
\hat\nu=2 (n_{s}+s+p)\,,\quad n_{s}=0,1,2,...
\end{equation}
One can show that the corresponding energy eigenfunctions  are square 
integrable only when $2p=-j$ and the quantum main number $n=2n_{s}+2s$ takes 
the values $n=0,1,...,n_{max}<j$.  Therefore, these are 
\begin{equation}\label{(solRM1)}
U_{j,n}(x)=N_{j,n}\sinh^{2s}\hat\omega x\cosh^{-j}\hat\omega x
F\left(-n_{s},
n_{s}-j+2s, 2s+\frac{1}{2}, -\sinh^{2}\hat\omega x\right)\,.
\end{equation}
Hence it results that the discrete energy spectrum is finite having 
$n_{max}+1$ levels where 
$n_{max}$ is the highest integer smaller than  $j$. This spectrum 
is included in the domain $[m, m\sqrt{1+1/\epsilon^{2}})$ since the energy 
levels are  
\begin{equation}\label{(labrm)}
{E_{j,n}}^{2}=\hat\omega^{2}[-(n-j)^{2}+(\epsilon^{2}+1)j(j+1)], \quad 
n=0,1,2... n_{max}\,.
\end{equation}  
The definitive form of the normalized energy eigenfunctions of the discrete 
spectrum will be calculated in the next section by using the shape invariance 
of the RM potentials.

The continuous spectrum cover the domain $[m\sqrt{1+1/\epsilon^{2}}, \infty)$ 
where we have $\hat \nu=i|\hat\nu|$. The generalized energy eigenfuctions of 
this spectrum are tempered distributions of the form (\ref{(solRM)}) with 
$2s=0,1$ and $2p=-j$. We note that for $m\to 0$ (when $j \to 0$) the discrete 
spectrum disappears while the continuous one becomes $[0, \infty)$. In this 
model the massless test particle move like in flat space-time and, therefore, 
it does not have nonrelativistic limit.

\subsection{The normal RO  and the nonrelativistic limit}
\

Our family of RO is continuous in $\lambda=0$ \cite{CO1}. This means that the 
limits for $\epsilon \to 0$ of the PT and RM models must coincide.  Indeed, 
according to Eqs.(\ref{(xPT)}) and (\ref{(xRM)}) we find that in this limit 
$x\to u$ while from Eqs.(\ref{(kpt)}) and (\ref{(krm)}) it results that for 
any model with $m\not=0$ we have $k \to \infty$, $j\to \infty$ but
\begin{equation}\label{(limk)}
\lim_{\epsilon \to 0}\epsilon^{2}k=
\lim_{\epsilon \to 0}\epsilon^{2}j=\frac{m}{\omega}\,.
\end{equation}
Furthermore, we can verify that the finite discrete spectra of the models with 
$\lambda>0$  become countable  while the continuous spectra disappear in a such 
a manner that the RM models and the PT ones have the same limit which is just 
the normal RO (with $\lambda=0$).
The special frame of this model coincides with the proper one  where 
the metric is defined by Eq.(\ref{(mno)}). Therefore, the  
relativistic potential is 
\begin{equation}\label{(vzero)}
V_{0}(u)=\lim_{\epsilon\to 0} V_{PT}(x)=  
\lim_{\epsilon\to 0} V_{RM}(x)=m^{2}\omega^{2}u^{2}  
\end{equation}
and the Klein-Gordon equation
\begin{equation}
\left[-\frac{d^2}{du^2}+m^{2}\omega^{2}u^{2}\right]U^{(0)}_{n}(u)=
({E_{n}}^{2}-m^{2})U^{(0)}_{n}(u)
\end{equation}
gives  the  familiar energy eigenfunctions of the NRHO, 
\begin{equation}
U_{n}^{(0)} 
=\left(\frac{m\omega}{\pi}\right)^{1/4}\frac{1}{\sqrt{n!2^{n}}}
e^{-m\omega^{2}u^{2}/2}H_{n}(\sqrt{m\omega}\,u)\,,   
\end{equation}
(where $H_{n}$ are  Hermite polynomials), but relativistic  energy levels,
\begin{equation}
{E_{n}^{(0)}}^{2}=m^{2}+2m\omega (n+\frac{1}{2})\,. 
\end{equation}

In the nonrelativistic limit (defined by $m/\omega \rightarrow \infty$), 
the normal RO becomes the NRHO with the potential $V_{0}/2m$ and usual energy 
levels. The nonrelativistic limit of the other models, with $m\not =0$ and 
$\lambda\not=0$,  can be easily calculated if we  observe that, according to 
Eqs.(\ref{(kpt)}) and (\ref{(krm)}), this is equivalent with the limit $\lambda 
\rightarrow 0$ and, in addition, $m \gg \omega$. Hereby it results that all the 
RO with $m>0$ have the same nonrelativistic limit like that of the normal RO of 
the mass $m$, namely the usual NRHO. On the other hand, it is interesting that 
in this way we can show that the parameter $\lambda$, or the parameter 
$\epsilon$ related to it, does not have a direct nonrelativistic equivalent, 
since  the terms involving $\lambda$ vanish in this limit. 
\newpage

\section{Supersymmetry and Shape invariance}
\    

A relativistic supersymmetric quantum mechanics can be constructed in the same 
way as the nonrelativistic one. The main problem here is to find the 
operator which should play the role of Hamiltonian. We shall show that   
this is the operator (\ref{(Delta)}) with a suitable translated 
potential.

\subsection{Supersymmetry}
\

Let us start with a (1+1)-dimensional relativistic model with the potential 
$V_{R}$ giving finite or countable energy spectrum. First we denote the energy 
levels by $E_{n}^{(-)}$ and the corresponding energy eigenfunctions by 
$U^{(-)}_{n}$. Then Eq.(\ref{(kkg)}) in the special frame can be written as
\begin{equation}\label{(min)}
{\bf \Delta}[V_{-}]U_{n}^{(-)}=d^{(-)}_{n}U_{n}^{(-)}, \qquad n=0,1,2,...
\end{equation}
where
\begin{equation}\label{(vminus)}
V_{-}=V_{R}-({E_{0}^{(-)}}^2-m^{2})
\end{equation}
and   
\begin{equation}\label{(dif)}
d_{n}^{(-)}={E_{n}^{(-)}}^{2}-{E_{0}^{(-)}}^{2}\,.
\end{equation}
Herein we have translated the spectrum of ${\bf \Delta}$ in a such a 
manner to accomplish the condition  
$d_{0}^{(-)}=0$  we need for defining the superpotential \cite{SUP} 
\begin{equation}\label{(sup)}
W(x)=-\frac{1}{U_{0}^{(-)}}\frac{dU_{0}^{(-)}(x)}{dx}\,. 
\end{equation}
Then we have $V_{-}=W^{2}-W'$ (with the notation $'=\partial_{x}$) and the  
supersymmetric partner (superpartner)  potential of $V_{-}$ reads  
$V_{+}=W^{2}+W'=-V_{-}+2W^{2}$. Furthermore,  we introduce the operator 
\begin{equation}\label{(opa)}
{\bf A}=-i{\bf P}+W({\bf X})  
\end{equation}
which satisfies 
\begin{equation}\label{(com)}
[{\bf A},{\bf A^{\dagger}}]=2W'({\bf X})
\end{equation}
and help us to write       
\begin{equation}\label{(aaa)}
{\bf \Delta}[V_{-}]={\bf A}^{\dagger}{\bf A}\,, \qquad
{\bf \Delta}[V_{+}]={\bf A}{\bf A}^{\dagger}\,.
\end{equation}

Now, like in the nonrelativistic case \cite{SUP}, we can convince ourselves 
that the spectrum of the eigenvalue problem 
\begin{equation}
{\bf \Delta}[V_{+}]U^{(+)}_{n}=d^{(+)}_{n}U^{(+)}_{n} 
\end{equation}
coincides with that of Eq.(\ref{(min)}), apart of the eigenvalue 
$d^{(-)}_{0}=0$. Thus we have  $d^{(+)}_{n}=d^{(-)}_{n+1}$, $n=0,1,2,..$,
while the normalized eigenfunctions of 
${\bf \Delta}[V_{-}]$ and ${\bf \Delta}[V_{+}]$ satisfy
\begin{equation}
{\bf A}U^{(-)}_{n}=\eta\sqrt{d^{(-)}_{n}} U^{(+)}_{n-1}\,, \qquad
{\bf A}^{\dagger}U^{(+)}_{n-1}=\eta^{*}\sqrt{d^{(-)}_{n}} U^{(-)}_{n}
\end{equation}
where $\eta$ is an arbitrary phase factor. 

Hence, we can say that the (1+1) relativistic supersymmetric quantum mechanics  
has the same main features as the nonrelativistic one. It remains us to study 
the shape invariance of the  relativistic potentials of the RO related through 
supersymmetry.

\subsection{Shape invariance}
\
  
Let us consider the PT model (k) and identify $U^{(-)}_{n}\equiv U_{k,n}$ and 
$E^{(-)}_{n}\equiv E_{k,n}$. Then the differences (\ref{(dif)}) are
\begin{equation}
d_{n}^{(-)}\equiv d_{k,n}={E_{k,n}}^{2}-{E_{k,0}}^{2}=\hat\omega^{2} n(n+2k),
\end{equation}
and from Eqs.(\ref{(vminus)}) and (\ref{(vpt)}) we obtain 
\begin{equation}
V_{-}(k,x)=V_{PT}(k,x)+m^{2}-{E_{k,0}}^2=\hat\omega^{2}[k(k-1)\tan^{2}
\hat\omega x -k],
\end{equation}
On the other hand, the normalized  ground-state eigenfunction  calculated 
from Eq.(\ref{(solPT1)}),   
\begin{equation}\label{(egpt)}
U_{k,0}(x)=\left(\frac{\hat\omega^2}{\pi}\right)^{\frac{1}{4}}\left[
\frac{\Gamma(k+1)}{\Gamma(k+\frac{1}{2})}\right]^{\frac{1}{2}}
\cos^{k}\hat\omega x \,,
\end{equation}
gives the superpotential $W(k,x)=\hat\omega k\tan\hat\omega x$ which allows us 
to find the superpartner of $V_{-}$,
\begin{equation}
V_{+}(k,x)=-V_{-}(k,x)+2W(k,x)^{2}=\hat\omega^{2}[k(k+1)\tan^{2}\hat\omega x 
+k]\,.
\end{equation}
Moreover, with this superpotential the operator (\ref{(opa)}) reads 
\begin{equation}\label{(opaPT)}
{\bf A}_{k}= -i{\bf P}+\hat\omega k\tan\hat\omega{\bf X}
=-\cos^{k}\hat\omega {\bf X}\, (i{\bf P})
\cos^{-k}\hat\omega {\bf X} 
\end{equation}
while from Eq.(\ref{(com)}) we obtain 
\begin{equation}
[{\bf A}_{k},{\bf A}_{k}^{\dagger}]=2k\hat\omega^{2}{\bf 1}
+\frac{1}{2k}({\bf A}_{k}+{\bf A}_{k}^{\dagger})^2\,.
\end{equation}

Now we  observe  that the potentials $V_{-}(k)$ and $V_{+}(k)$ are 
shape invariant since
\begin{equation}\label{(sh)}
V_{+}(k,x)=V_{-}(k+1,x)+\hat\omega^{2} (2k+1).
\end{equation}
Consequently, we can identify $U_{n}^{(+)}\equiv U_{k+1,n}$ which means that  
the normalized energy eigenfunctions satisfy
\begin{equation}\label{(aapt)}
{\bf A}_{k}U_{k,n}=\sqrt{d_{k,n}}U_{k+1,n-1}, \qquad
{\bf A}_{k}^{\dagger}U_{k+1,n-1}=\sqrt{d_{k,n}}U_{k,n}. 
\end{equation}
as it results from Eqs.(\ref{(aaa)}) with $\eta=1$.
Thus we have related the energy eigenfunctions of the model $(k)$ with those of 
its superpartner model, $(k+1)$. In general, we can write any normalized 
energy eigenfunction of the model $(k)$ as
\begin{equation}\label{(RPT)}
U_{k,n}=\frac{1}{\hat\omega^{n}\sqrt{n!}}\left[\frac{\Gamma(n+2k)}
{\Gamma(2n+2k)}\right]^{\frac{1}{2}}
{\bf A}_{k}^{\dagger}{\bf A}_{k+1}^{\dagger}...{\bf A}_{k+n-1}^{\dagger}
U_{k+n,0}.
\end{equation}
where $U_{k+n,0}$ is the normalized ground-state eigenfunction of the model 
$(k+n)$ given by Eq.(\ref{(egpt)}).

For the relativistic RM models we use the same  method 
starting with the model $(j)$ and denoting  
$U_{n}^{(-)}\equiv U_{j,n}$ and $E_{n}^{(-)}\equiv E_{j,n}$. 
Then the differences (\ref{(dif)}) are
\begin{equation}
d^{(-)}_{n}\equiv d_{j,n}={E_{j,n}}^{2}-{E_{j,0}}^{2}=\hat\omega^{2}n(2j
-n),
\end{equation}
and, according to (\ref{(vrm)}), we have 
\begin{equation}
V_{-}(j,x)=V_{RM}(j,x)+m^{2}-E_{j,0}^2
=\hat\omega^{2}[j(j+1)\tanh^{2}\hat\omega x -j].
\end{equation}
From Eq.(\ref{(solRM1)}) we find the normalized  ground-state eigenfunction 
\begin{equation}\label{(egrm)}
U_{j,0}(x)=\left(\frac{\hat\omega^2}{\pi}\right)^{\frac{1}{4}}\left[
\frac{\Gamma(j+\frac{1}{2})}{\Gamma(j)}\right]^{\frac{1}{2}}
\cosh^{-j}\hat\omega x\,,
\end{equation}
giving the superpotential $W(j,x)=\hat\omega j\tanh\hat\omega x$.  Hereby we 
obtain 
\begin{equation}
V_{+}(j,x)=-V_{-}(j,x)+2W(j,x)^{2}=
\hat\omega^{2}[j(j-1)\tanh^{2}\hat\omega x +j]\,.
\end{equation}
Now the operator (\ref{(opa)}) reads 
\begin{equation}\label{(opaRM)}
{\bf A}_{j}= -i{\bf P}+\hat\omega j\tanh\hat\omega{\bf X}
=-\cosh^{-j}\hat\omega {\bf X}\, (i{\bf P})
\cosh^{j}\hat\omega {\bf X} 
\end{equation}
while Eq.(\ref{(com)}) gives
\begin{equation}
[{\bf A}_{j},{\bf A}_{j}^{\dagger}]=2j\hat\omega^{2}{\bf 1}-
\frac{1}{2j}({\bf A}_{j}+{\bf A}_{j}^{\dagger})^2.
\end{equation}

The potentials $V_{-}(j)$ and $V_{+}(j)$ are shape invariant since
\begin{equation}\label{(sh)}
V_{+}(j,x)=V_{-}(j-1,x)+\hat\omega^{2}(2j-1).
\end{equation}
Consequently, as in the previous case, we find that the normalized energy 
eigenfunctions satisfy
\begin{equation}\label{(aa)}
{\bf A}_{j}U_{j,n}=\sqrt{d_{j,n}}U_{j-1,n-1}, \qquad
{\bf A}_{j}^{\dagger}U_{j-1,n-1}=\sqrt{d_{j,n}}U_{j,n} 
\end{equation}
if we take $\eta=1$ in Eqs.(\ref{(aaa)}). Thus we have obtained the relation 
between the sets of energy eigenfunctions of the superpartner models $(j)$ and 
$(j-1)$. Moreover, we can also express  the normalized eigenfunctions as  
\begin{equation}\label{(RRM)}
U_{j,n}=\frac{1}{\hat\omega^{n}\sqrt{n!}}\left[\frac{\Gamma(2j-2n+1)}
{\Gamma(2j-n+1)}
\right]^{\frac{1}{2}}
{\bf A}_{j}^{\dagger}{\bf A}_{j-1}^{\dagger}...{\bf A}_{j-n+1}^{\dagger}
U_{j-n,0}.
\end{equation}
where now $U_{j-n,0}$ is the normalized ground-state eigenfunction of the 
model $(j-n)$ given by Eq.(\ref{(egrm)}).

\subsection{The normalized energy eigenfunctions}
\

The  normalization of the energy eigenfunctions of the PT models may be easily 
done in usual way but for the RM models there are some technical difficulties 
that can be avoided  by  using the previous results. Indeed, we observe 
that the Eqs.(\ref{(RPT)}) and (\ref{(RRM)}) are nothing else than the operator 
form of the Rodrigues formulas of the normalized eigenfunctions (in our phase 
convention with $\eta=1$). Therefore, it remains  only to rewrite their 
expressions in usual form. 
 
For the  PT models  we replace the operator (\ref{(opaPT)})  
in Eq.(\ref{(RPT)}) which takes the form  
\begin{eqnarray}\label{(RPTf)}
U_{k,n}(x)=\frac{(-1)^n}{\hat\omega^{n}\sqrt{n!}}\left[\frac{\Gamma(n+2k)}
{\Gamma(2n+2k)}\right]^{\frac{1}{2}}
\cos^{-k}\hat\omega x\frac{d}{dx}\frac{1}{\cos\hat\omega x}
\frac{d}{dx}\cdots\nonumber\\   
\cdots\frac{1}{\cos\hat\omega x}\frac{d}{dx}\cos^{k+n-1}\hat\omega x   
U_{k+n,0}(x)\,.
\end{eqnarray}
Then, according to Eqs.(\ref{(xPT)}) and (\ref{(egpt)}), we obtain the final 
Rodrigues formula of the normalized energy eigenfunctions of the PT models 
in  proper frames, 
\begin{eqnarray}\label{(RPTf)}
U_{k,n}(u)=\left(\frac{\hat\omega^2}{\pi}\right)^{\frac{1}{4}}
\frac{(-1)^n}{\hat\omega^{n}\sqrt{n!}}\left[\frac{\Gamma(2k+n)\Gamma(k+n+1)}
{\Gamma(2k+2n)\Gamma(k+n+\frac{1}{2})}\right]^{\frac{1}{2}}\nonumber\\   
\times(1-\hat\omega^{2} u^{2})^{-\frac{k-1}{2}}\frac{d^{n}}{du^n}
(1-\hat\omega^{2} u^{2})^{k+n-\frac{1}{2}}\,.   
\end{eqnarray}

In the same way we can derive the Rodrigues formula for the normalized energy 
eigenfunctions of the RM models. By using the Eqs.(\ref{(opaRM)})
and (\ref{(RRM)})  we find  
the normalized energy egenfunctions of the discrete spectrum in the proper 
frame,  
\begin{eqnarray}\label{(RRMf)}
U_{j,n}(u)=\left(\frac{\hat\omega^2}{\pi}\right)^{\frac{1}{4}}
\frac{(-1)^n}{\hat\omega^{n}\sqrt{n!}}\left[\frac{\Gamma(2j-2n+1)
\Gamma(j-n+\frac{1}{2})}
{\Gamma(2j-n+1)\Gamma(j-n)}\right]^{\frac{1}{2}}\nonumber\\   
\times(1+\hat\omega^{2} u^{2})^{\frac{j+1}{2}}\frac{d^{n}}{du^n}
(1+\hat\omega^{2} u^{2})^{-j+n-\frac{1}{2}}\,.   
\end{eqnarray}
Of course, as it was expected, this formula gives square integrable functions 
only for $n\le n_{max}$. On the other hand, here the problem of "normalization" 
of the generalized energy eigenfuntions of the continuous spectrum remains open 
since there are not yet efficient procedures for doing this.

Hence we have obtained the definitive formulas of the normalized energy 
eigenfunctions of our RO corresponding to the discrete energy levels. These 
can be written now in terms of Jacobi or Gegenbauer polynomials \cite{AS} and 
even as associated Legendre functions \cite{PT1,PT2} but only when $k$ and 
$j$ are integer numbers. In the particular case of the PT model with $k=1$ the 
trigonometric form (\ref{(u1n)}) can be derived  from Eq.(\ref{(RPTf)}) 
by using the properties of the Tchebyshev polynomials. 

In Sec.3.3 we have seen that the normal RO has the same  energy eigenfunctions 
as the NRHO. Now, by taking into account 
that for large arguments, $z$, we have $\Gamma(z+a)/\Gamma(z+b)\sim z^{(a-b)}$ 
and by using Eqs.(\ref{(limk)}) we  verify directly that   
\begin{equation}
\lim_{\epsilon\to 0}U_{k,n}(u)=
\lim_{\epsilon\to 0}U_{j,n}(u)=U^{(0)}_{n}(u)\,.
\end{equation}
Because of these properties we can say that the eigenfunctions (\ref{(RPTf)}) 
and (\ref{(RRMf)}) represent   relativistic generalizations of the  
NRHO eigenfunctions, different from that of the algebraic method \cite{OSC}.


\section{Shift operators}
\

In our models  only one main quantum number is involved and, therefore, 
in each model we must have a pair of shift operators, i.e. the rising and the 
lowering operators of the energy basis. In general, the shift operators are 
different from those of the supersymmetry apart the shift operators of the 
normal RO that are up to factors just those of the supersymmetry since
this model is its own superpartner.  

Let us start with this simplest case since here the energy eigenfunctions are 
similar to those of the NRHO. Consequently, we can take over the well-known 
results from the nonrelativistic theory defining the differential 
operators  
\begin{eqnarray}                                                                                                  
({\bf a}U)(u)&=& {1 \over \sqrt{2 m \omega}} \left({d \over d u} + 
m \omega u \right) U(u),\\
({\bf a}^{\dagger}U)(u)&=&{1\over \sqrt{2 m \omega}} \left( - {d \over d u} + 
m \omega u \right)U(u).
\end{eqnarray}
of the Heisenberg-Weyl algebra. Obviously, they are the desired shift operators 
which obey  $[{\bf a}, {\bf a}^{\dagger}]={\bf 1}$  giving us the operator of 
number of quanta  ${\bf N}= {\bf a}^{\dagger} {\bf a}$ and  
\begin{equation}
{\bf X}= \frac{1}{\sqrt{2m\omega}}({\bf a}^{\dagger}+{\bf a})\,, \quad
{\bf P}=-i\sqrt{\frac{m\omega}{2}} ({\bf a}^{\dagger}-{\bf a})\,.
\end{equation}
Moreover, it is natural to find that     
\begin{equation}
\lim_{\epsilon\to 0}{\bf A}_{k} = 
\lim_{\epsilon\to 0}{\bf A}_{j}= \sqrt{2m\omega}\,{\bf a}\,.
\end{equation}  

For the models with  $\lambda\not= 0$ the shift operators  differ from those 
of supersymmetry. They can be calculated directly by using the action of the 
supesymmetry operators and  the form of the normalized energy 
eigenfunctions  derived above.  
In the case of  $\lambda=-\epsilon^{2}$, after a few 
manipulation, we find that the shift operators of the PT model $(k)$ can be 
defined in the proper frame as  
\begin{eqnarray}
({\bf A}_{k,(+)}U_{k,n})(u)&=&\frac{1}{\hat\omega\sqrt{2k}} 
\left[-(1-\hat\omega^{2}u^{2}) {d\over du} +
 \hat\omega^2 u ( k + n ) \right]U_{k,n}(u)\,,\label{(apPT)} \\
({\bf A}_{k,(-)}U_{k,n})(u)&=&\frac{1}{\hat\omega\sqrt{2k}} 
\left[(1-\hat\omega^{2}u^{2}) {d\over du} +
 \hat\omega^2 u ( k + n ) \right] U_{k,n}(u)\,.\label{(amPT)}
\end{eqnarray}                                                
Their shifting action is  
\begin{equation}
{\bf A}_{k,(+)}U_{k,n}=C_{k,n}^{(+)}U_{k,n+1}\,,\quad
{\bf A}_{k,(-)}U_{k,n}=C_{k,n}^{(-)}U_{k,n-1}\,,
\end{equation}
where 
\begin{eqnarray}
C_{k,n}^{(+)}&=& {1\over  \sqrt{2k}} \, \left[ {(2 k + n)( k + n)
 \over  k + n + 1 }\right]^{1\over 2} \sqrt{n+1}\,,\label{(3250)}\\                                                 
C_{k,n}^{(-)}&=& {1\over  \sqrt{2k}} \, \left[ {(2 k + n - 1)( k + n)
 \over  k + n - 1 }\right]^{1\over 2} \sqrt{n}\,.\label{(3260)}
\end{eqnarray}
If we rewrite the action of the operators (\ref{(apPT)}) and (\ref{(amPT)}) in 
the special frame $(t,x)$ then we recover the result of Ref.\cite{PT1}.  
Furthermore, we can verify the commutation relation 
\begin{equation}\label{(3290)}
[{\bf A}_{k,(-)},{\bf A}_{k,(+)}]U_{k,n}=\left(1+\frac{n}{k}\right)U_{k,n}
\end{equation}
and the identity  
 \begin{equation}\label{(3291)}
2k{\bf A}_{k,(+)}{\bf A}_{k,(-)}U_{k,n}=n(2k+n-1)U_{k,n}
\end{equation}
which is just the Klein-Gordon equation in operator form \cite{CA2}. 
In the limit $\epsilon\to 0$ we have \cite{CA1}
\begin{equation}
\lim_{\epsilon \to 0}{\bf A}_{k,(+)}={\bf a}^{\dagger}\,,\quad   
\lim_{\epsilon \to 0}{\bf A}_{k,(-)}={\bf a}\,. 
\end{equation}   

With the same procedure we find  the shift operators of the RM 
model $(j)$ in the proper frame, 
\begin{eqnarray}
({\bf A}_{j,(+)}U_{j,n})(u)&=&\frac{1}{\hat\omega\sqrt{2j}} 
\left[-(1+\hat\omega^{2}u^{2}) {d\over du} +
 \hat\omega^2 u ( j - n ) \right]U_{j,n}(u)\,, \\
({\bf A}_{j,(-)}U_{j,n})(u)&=&\frac{1}{\hat\omega\sqrt{2j}} 
\left[(1+\hat\omega^{2}u^{2}) {d\over du} +
 \hat\omega^2 u ( j - n ) \right] U_{j,n}(u)\,,
\end{eqnarray}                                                
which have the action 
\begin{equation}
{\bf A}_{j,(+)}U_{j,n}=C_{j,n}^{(+)}U_{j,n+1}\,,\quad 
{\bf A}_{j,(-)}U_{j,n}=C_{j,n}^{(-)}U_{j,n-1}\,,
\end{equation}
where 
\begin{eqnarray}
C_{j,n}^{(+)}&=& {1\over  \sqrt{2j}} \, \left[ {(2 j - n)( j - n)
 \over  j - n - 1 }\right]^{1\over 2} \sqrt{n+1}\,,\label{(32501)}\\                                                 
C_{j,n}^{(-)}&=& {1\over  \sqrt{2j}} \, \left[ {(2 j - n + 1)( j - n)
 \over  j - n + 1 }\right]^{1\over 2} \sqrt{n}\,.\label{(32601)}
\end{eqnarray}
They obey the commutation rule  
\begin{equation}\label{(32901)}
[{\bf A}_{j,(-)},{\bf A}_{j,(+)}]U_{j,n}=\left(1-
\frac{n}{j}\right)U_{j,n}
\end{equation}
and give us  the operator form of the Klein-Gordon equation for discrete 
levels, 
 \begin{equation}\label{(32911)}
2j{\bf A}_{j,(+)}{\bf A}_{j,(-)}U_{j,n}=
n(2j-n+1)U_{j,n}\,.
\end{equation}
Finally we find that  
\begin{equation}
\lim_{\epsilon \to 0}{\bf A}_{j,(+)}={\bf a}^{\dagger}\,,\quad   
\lim_{\epsilon \to 0}{\bf A}_{j,(-)}={\bf a}\,. 
\end{equation}   

We must specify that the shift operators of the models with $\lambda \not =0$ 
have two important properties, namely: they are not pure differential operators 
on ${\cal L}^{2}(D_{u},\mu)$ and, moreover, the raising and  lowering 
operators are not adjoint with each other, i.e. ${\bf A}_{k,(\pm)}\not=
({\bf A}_{k,(\mp)})^{\dagger}$ and similarly for the RM models.

\section{Comments}
\

In this article we have studied the quantum modes of a family of (1+1) RO by 
using the methods of a supersymmetric relativistic quantum mechanics 
similar with the well-known nonrelativistic one. This was possible since the 
form of the Klein-Gordon equation in the special frames is very close to that 
of the Schr\" odinger equation,  allowing us to introduce the 
relativistic potentials involved in supersymmetry and to exploit their shape 
invariance. 

However,  our relativistic theory has  some new interesting features due to 
the fact that the mass is not only involved in the formula of the energy levels 
but also plays here  the role of a coupling constant. For this reason there are  
some kind of regularities leading to a very simple parametrization  
in a such a manner that for any pair of superpartner models  we have 
either $\Delta k=\pm 1$ or $\Delta j=\mp1$. Thus $k$ and $j$ simulate the 
behavior of quantum numbers even though they can not be considered as 
eigenvalues of self-adjoint operators \cite{CA1}. On the other hand, the models 
with superpartner potentials can be seen as having particles of different 
masses moving on the same background. The consequence is that the masses of the 
sets of superpartner PT or RM models  appear as being quantized according to 
the formulas $m_{k}^{2}=\epsilon^{2}\hat\omega^{2}k(k-1)$ and 
 $m_{j}^{2}=\epsilon^{2}\hat\omega^{2}j(j+1)$  respectively. These remarkable 
properties  helped us to easily write down the Rodrigues formulas of the 
normalized energy eigenfunctions of the discrete spectra and to find the 
corresponding shift operators.

Concluding we can say that our family of models brings together the main 
solvable problems with parity-symmetric potentials of the one-dimensional 
quantum  mechanics, interpreted as relativistic oscillators in the sense that 
all these models (apart those with $k=1$ and $j=0$) lead to the NRHO in the 
nonrelativistic limit.

\end{document}